\documentclass[review]{elsarticle}

\usepackage{lineno,hyperref}
\usepackage{amsmath, amssymb, amsfonts}
\modulolinenumbers[5]
\usepackage{graphicx}
\usepackage{caption}
\usepackage{subcaption}
\usepackage{natbib}
\usepackage{threeparttable}
\usepackage{xcolor}

\usepackage{amsthm}

\theoremstyle{definition}

\newcommand{\nbar}{\bar{\mathbf n}}
\newcommand{\sbar}{\bar{\mathbf s}}
\newcommand{\wbar}{\bar{\mathbf w}}
\newcommand{\Cbar}{\bar{\mathbf C}}
\newcommand{\half}{\frac{1}{2}}
\newcommand{\quarter}{\frac{1}{4}}

\journal{Journal of \LaTeX\ Templates}
\usepackage[top=1in, left=1in, right=1in, bottom=1in]{geometry}

\begin{document}

\begin{frontmatter}

\title{Derived Crystal Structure of Martensitic Materials by Solid-Solid Phase Transformation}

\author[1]{Mostafa Karami}
\author[2]{Nobumichi Tamura}
\author[3]{Yong Yang}
\author[1]{Xian Chen\corref{corr}}


\address[1]{Mechanical and Aerospace Engineering, Hong Kong University of Science and Technology, Hong Kong}
\address[2]{Advanced Light Source, Lawrence Berkeley National Laboratory, California, United States}
\address[3]{Mechanical Engineering, City University of Hong Kong, Hong Kong}
\cortext[corr]{Corresponding author}
\ead{xianchen@ust.hk}

\begin{abstract}
We propose a mathematical description of crystal structure: underlying translational periodicity together with the distinct atomic positions up to the symmetry operations in the unit cell. It is consistent with the international table of crystallography. By the Cauchy-Born hypothesis, such a description can be integrated with the theory of continuum mechanics to calculate a \emph{derived crystal structure} produced by solid-solid phase transformation. In addition, we generalize the expressions for orientation relationship between the parent lattice and the derived lattice. The derived structure rationalizes the lattice parameters and the general equivalent atomic positions that assist the indexing process of X-ray diffraction analysis for low symmetry martensitic materials undergoing phase transformation. The analysis is demonstrated in a CuAlMn shape memory alloy. From its austenite phase (L$2_1$ face-centered cubic structure), we identify that the derived martensitic structure has the orthorhombic symmetry P$mmm$ with derived lattice parameters $a_\text{dv} = 4.36491 \AA$, $b_\text{dv} = 5.40865 \AA$ and $c_\text{dv} = 4.2402 \AA$, by which the complicated X-ray Laue diffraction pattern can be well indexed, and the orientation relationship can be verified.           
\end{abstract}


\end{frontmatter}

\section{Introduction}

Materials undergoing reversible martensitic transformations show great potentials in many emerging applications such as biomedical implants, nano/microactuators and solid state caloric coolings. The underlying functionality of using these materials is the ability to recover a large macroscopic deformation (i.e. 5\% -- 10\%) \cite{chang1951, tadaki1998cu, miyazaki1982characteristics} during the reversible structural transformation. Many applications require these materials to run millions of transformation cycles, their functionality typically degrades quickly in the first couple of hundreds of cycles, even for the most successful commercial alloy -- Nitinol. \cite{otsuka2005physical} 
Recent significant advances in developing ultra-low fatigue martensitic materials \cite{Song_2013, chluba_2015} show that the design of phase-transforming materials can be guided through some kinematic compatibility principles called cofactor conditions \cite{Chen_2013}, i.e. the super compatibility conditions for the existence of stressed-free microstructure during phase transformation. When the cofactor conditions are satisfied, the thermal hysteresis is minimized without compromising the amount of latent heat. \cite{Song_2013} Meanwhile the thermomechanical response does not degrade at all even upon tens of million mechanically induced transformation cycles. \cite{chluba_2015} These discoveries underlie a theory-driven design strategy for phase transforming materials, of which the most crucial step is to precisely determine the crystal structures of the transforming material as austenite (high symmetry structure in high temperature phase) and martensite (low symmetry structure in low temperature phase). However, this step is non-trivial, and often quite tedious. 

In principle, the structural parameters of a crystalline solid are determined by X-ray diffraction (XRD) experiments. One of the most common XRD measurement used for structural determination is Rietveld refinement of powder diffraction data obtained with either CuK$\alpha$ or MoK$\alpha$ radiation \cite{young1993rietveld}. The testing specimen should be either in powder or bulk form with sufficient randomization of grain orientations illuminated by the monochromatic X-ray beam. However, the as-cast metallic specimen after proper heat treatment produced in laboratory is mostly in bulk form with coarse grain size. For example, the grain size of common Cu-based $\beta$ alloys is about 200 -- 500 $\mu$m. \cite{bhattacharya2003}  The orientation randomness of the specimen from the lab production is insufficient for ordinary XRD powder method, especially for the low symmetry structures. In most cases, the crystal structure of the developed material is unknown, which makes the Rietveld analysis impossible. Structure solution through single crystal x-ray diffraction requires isolated good quality single crystals and is therefore hardly applicable to the bulk samples. The lack of structural knowledge for low symmetry metallic materials highly hinders the material development for desirable properties. Therefore, it is very important to have a unified way for structural determination. 

For martensitic materials, most are inherited from a high symmetry phase, austenite, through solid-solid phase transformation \cite{Otsuka_1999}. The formation of martensite microstructure is strongly restricted by the crystallographic compatibility to the austenite structure. Therefore, the nature of the phase transformation and the orientation relationships between austenite and martensite can be used to propose a universal structural determination method for the martensite crystal structure transformed from the cubic austenite. 
The cubic structures, including simple cubic, face-centered cubic and body-centered cubic, have only one structural dimensional parameter, $a_0$ that can be accurately determined by ordinary XRD experiments, from which we can derive the primitive lattice metric -- the underlying 3-dimensional periodicities of the martensite lattice -- through the crystallographic relationships with the austenite phase and the Cauchy-Born rule for solid-solid phase transformation. Such a derivation does not require any pre-knowledge of the crystal symmetry or lattice parameters. The derived lattice metric can be slightly perturbed from the reference lattice and used to present the unknown martensite structure for advanced structural characterization and analysis such as XRD, EBSD and so on. In this paper, we lay down the fundamental formulation for the derived lattice from a cubic structure through the solid-solid phase transformation. We then used synchrotron Laue x-ray microdiffraction experiment combined with energy scans to demonstrate our method for an unknown Cu-based $\beta$ alloy. To bridging the discrepancy between the previous mathematical description of lattice \cite{ball1992, pitteri_1998symm} and the symmetry calculation of crystal structures by the international union of crystallography \cite{hahn1983}, we propose a modified description for the \emph{crystal structure} consisting of two parts: underlying translational periodicity and the fractional atomic positions in the unit cell in consistent with the general equivalent positions used in the international table of crystallography. Consider the parent and child phases as two discrete vector spaces mapped by a homogeneous linear transformation, the orientation relationship can be expressed by the \emph{lattice correspondence matrix}, by which specific lattice vector/plane parallelisms can be derived.  

\section{Derived lattice from solid-solid phase transformation}

\subsection{Mathematical representation of lattice and crystal structure}
For any three-dimensional Bravais lattice  \cite{ball1992}:
\begin{equation} \label{eq:bravais}
\mathcal L(\mathbf E) = \{\mathbf E \mathbf n: \forall \mathbf n \in \mathbb Z^{3}\},
\end{equation} 
the invertible matrix $\mathbf E \in \mathbb R^{3 \times 3}$ is the lattice basis. Its three linearly independent column vectors are the \emph{lattice vectors}. The symmetry of a Bravais lattice $\mathcal L (\mathbf E)$ is represented by an orthogonal tensor $\mathbf R \in O(3)$ such that :
\begin{equation}
\mathbf R \mathbf E = \mathbf E \mathbf L
\end{equation}
for an integral matrix $\mathbf L \in GL(3)$. All symmetry operations of the Bravais lattice $\mathcal L(\mathbf E)$ form a finite group $\mathcal P$, defined as the point group of $\mathcal L(\mathbf E)$. \cite{ball1992, pitteri_1998symm} To express the lattice parameters, we introduce the \emph{lattice metric} tensor \cite{pitteri_1998symm}:
\begin{equation}\label{eq:lm}
\mathbf C = \mathbf E^T \mathbf E \text{ for a Bravais lattice } \mathcal L(\mathbf E).
\end{equation}
The lattice metric tensor is always positive definite and symmetric. The lattice parameters of $\mathcal L(\mathbf E)$ are a sextuplet $\mathbf p = (p_1, p_2, p_3, p_4, p_5, p_6)$ depending on the lattice metric tensor. $p_i = \sqrt{\mathbf C_{ii}}$ (no sum) for $i = 1, 2, 3$, and $p_4, p_5, p_6 \in [-1, 1]$ satisfy:
\begin{equation}\label{eq:lp}
p_4 = \frac{\mathbf C_{2, 3}}{p_2 p_3}, ~ p_5 = \frac{\mathbf C_{1, 3}}{p_1 p_3}, p_6 = \frac{\mathbf C_{1, 2}}{p_1 p_2}.
\end{equation}

\begin{table}[ht]
\caption{Lattice parameters sextuplets of the 14 Bravais lattices}\label{tb:lp}
\begin{threeparttable}
\begin{tabular}{c|c|c|c}
& Bravais lattice & Lattice parameters & Order of the point group\tnote{1}\\ \hline
1 & simple cubic (sc) & $a_0(1, 1, 1, 0, 0, 0)$ & 24 \\ \hline
2 & face-centered cubic (fcc) & $(\frac{a_0}{\sqrt{2}}, \frac{a_0}{\sqrt{2}}, \frac{a_0}{\sqrt{2}}, \frac{1}{2}, \frac{1}{2}, \frac{1}{2})$ & 24 \\ \hline
3 & body-centered cubic (bcc) & $(\frac{\sqrt{3}a_0}{2}, \frac{\sqrt{3}a_0}{2}, \frac{\sqrt{3}a_0}{2}, \frac{1}{3}, \frac{1}{3}, -\frac{1}{3})$ & 24 \\ \hline
4 & hexagonal & $(a, a, c, 0, 0, \frac{1}{2})$ & 12 \\ \hline
5 & trigonal & $(a, a, a, \cos \alpha, \cos \alpha, \cos \alpha)$ & 6 \\ \hline
6 & tetragonal & $(a, a, c, 0, 0, 0)$ & 8 \\ \hline
7 & body-centered tetragonal (bct) & $(\alpha, \alpha, \alpha, \frac{-2a^2+c^2}{4\alpha^2}, \frac{c^2}{4\alpha^2}, \frac{c^2}{4\alpha^2})\tnote{2}$ & 8 \\ \hline
8 & primitive orthorhombic (po) & $(a, b, c, 0, 0, 0)$ & 4 \\ \hline
9 & base-centered orthorhombic (bco) & $(\frac{\gamma}{2}, \frac{\gamma}{2}, c, 0, 0, \frac{-a^2 + b^2}{\gamma^2})\tnote{3}$ & 4 \\ \hline
10 & face-centered orthorhombic (fco) & $(\frac{\gamma}{2}, \frac{\alpha}{2}, \frac{\beta}{2}, \frac{c^2}{\beta \alpha}, \frac{a^2}{\gamma \beta}, \frac{b^2}{\gamma \alpha})\tnote{3}$ & 4 \\ \hline
11 & body-centered orthorhombic (ico) & $(\frac{\alpha}{2}, \frac{\alpha}{2}, \frac{\alpha}{2}, \frac{\alpha^2 - 2 b^2}{\alpha^2}, \frac{-\alpha^2 + 2 c^2}{\alpha^2}, \frac{\alpha^2 - 2 a^2}{\alpha^2})\tnote{4}$ & 4 \\ \hline
12 & primitive monoclinic (pm) & $(a, b, c, 0, \cos\beta, 0)$ & 2 \\ \hline
13 & base-centered monoclinic (bcm) & $(\frac{m}{2}, \frac{m}{2}, c, - \frac{a c \cos \beta}{m c}, \frac{a c \cos \beta}{m c}, \frac{-a^2 + b^2}{m^2})\tnote{5}$ & 2 \\ \hline
14 & triclinic & $(a, b, c, \cos\alpha, \cos \beta, \cos \gamma)$ & 1 \\ \hline
\end{tabular}
{\small
\begin{tablenotes}
\item[1] We only consider the proper rotational symmetry here.
\item[2] $\alpha = \frac{\sqrt{2a^2 + c^2}}{2}$
\item[3] $\alpha = \sqrt{b^2 + c^2}$, $\beta = \sqrt{a^2 + c^2}$, $\gamma = \sqrt{a^2 + b^2}$
\item[4] $\alpha = \sqrt{a^2 + b^2 + c^2}$
\item[5] $m = \sqrt{a^2 + b^2}$
\end{tablenotes}}
\end{threeparttable}
\end{table}

One can consider the underlying periodicity of a lattice by either the lattice metric tensor or the lattice parameters, which are both invariant under symmetry operations and rigid body rotations. Table \ref{tb:lp} lists the expressions of lattice parameters for all 14 Bravais lattices in 3-dimension written in their primitive basis, i.e. the basis underlies the smallest unit cell defined as $\mathcal U(\mathbf E) = \{\mathbf E \lambda: \forall~ (\lambda \cdot \lambda) \in [0, 1)\}$, which only consists of one lattice point. However, crystallography theory does not deal with the primitive lattice basis, because it is not always orthogonal for all types of Bravais lattices. To facilitate the crystallographic calculations, X-ray crystallographers use the \emph{conventional basis} in their formula and equations. There are 7 out of 14 Bravais lattices in Table \ref{tb:lp}: fcc, bcc, bct, bco, fco, ico and bcm, whose primitive basis is not consistent with their conventional basis. Therefore, the expression of lattice parameters in primitive basis written in Table \ref{tb:lp} for these Bravais lattices is different from what we usually use for crystallographic calculation.  

The conventional lattice described by the conventional basis is a sublattice of the original Bravais lattice. A sublattice can be considered as a multilattice defined as:
\begin{equation}\label{eq:mlattice}
\mathcal M(\mathbf E; \mathbf w_i) = \{\mathbf E (\mathbf n + \mathbf w_i): \forall \mathbf n \in \mathbb Z^3, \ \text{for some } \mathbf w_i \in \mathbb R^3 \text{ with } |\mathbf w_i| \in [0, 1), i = 1, 2, ..., m\}.
\end{equation}
Here, the basis $\mathbf E$ is not necessarily the primitive basis. By the definition of \eqref{eq:mlattice}, there exist $m$ lattice points in the unit cell $\mathcal U(\mathbf E)$. The fractional vectors $\mathbf w_i, i = 1, ..., m$ can be interpreted as $m$ Bravais lattices that are displaced by $\mathbf E \mathbf w_i$ respective to each other.  For example, we can choose $a_0 \mathbf I$ as the basis of a multilattice to express the face-centered cubic unit cell: 
\begin{equation}\label{eq:mfcc}
\renewcommand*{\arraystretch}{0.75}
\mathcal M_\text{fcc}(a_0 \mathbf I; \mathbf w_1, \mathbf w_2, \mathbf w_3, \mathbf w_4), \text{ for } \mathbf w_1 = \begin{bmatrix}0\\ 0\\ 0\end{bmatrix}, \ \mathbf w_2 = \begin{bmatrix}\frac{1}{2}\\ \frac{1}{2}\\ 0\end{bmatrix}, \ \mathbf w_3 = \begin{bmatrix} 0\\ \frac{1}{2}\\ \frac{1}{2}\end{bmatrix}, \ \mathbf w_4 = \begin{bmatrix} \frac{1}{2}\\ 0 \\ \frac{1}{2}\end{bmatrix}.
\renewcommand*{\arraystretch}{1}
\end{equation}
 It describes the same periodicity as what $\mathcal L(\mathbf E)$ does where $\mathbf E$ is the primitive lattice basis given by
\begin{equation}\label{eq:primitivefcc}
\renewcommand*{\arraystretch}{0.75}
\mathbf E = \frac{a_0}{2}\begin{bmatrix}1 & 0 &1\\1 & 1 & 0\\0& 1 & 1 \end{bmatrix} = a_0 \mathbf I \chi, \text{ where } \chi = (\mathbf w_2, \mathbf w_3, \mathbf w_4).
\renewcommand*{\arraystretch}{1}
\end{equation}
\begin{proof}
For any lattice point in $\mathcal M_\text{fcc}$, it is expressed as :
\[a_0\mathbf I (\mathbf n + \mathbf w_i) = \mathbf E \chi^{-1} (\mathbf n + \mathbf w_i) = \mathbf E (\tilde{\mathbf n} + \chi^{-1}\mathbf w_i) = \left\{\begin{array}{l}\mathbf E \tilde{\mathbf n} \text{ for } i = 1\\ 
\mathbf E (\tilde{\mathbf n} + [1, 0, 0]^T) \text{ for } i = 2 \\ 
\mathbf E (\tilde{\mathbf n} + [0, 1, 0]^T) \text{ for } i = 3\\
\mathbf E (\tilde{\mathbf n} + [0, 0, 1]^T) \text{ for } i = 4\end{array}\right.,
\]
for some $\tilde{\mathbf n}  = \chi^{-1}\mathbf n \in \mathbb Z^3$.
\end{proof}

Among the 14 Bravais lattices, there are 7 choices of the conventional basis corresponding to the 7 crystal systems. They are the bases of simple cubic, hexagonal, trigonal, primitive tetragonal, primitive orthorhombic, primitive monoclinic (with b-axis as the unique axis) and triclinic. The periodicity of the other non-primitive Bravais lattices can be expressed by the multilattice description using the corresponding conventional basis. Then the integer triplet $\mathbf n = [n_1, n_2, n_2]$ in \eqref{eq:mlattice} is consistent with the notation of Miller indices for crystallographic direction introduced by William H Miller in crystallography. Since the reciprocal basis is derived by taking the inverse of the transpose of the real lattice basis, all the calculations for the reciprocal lattice remains the same as given in \eqref{eq:mlattice}, except that the integer triplet $\mathbf n$ represents the index of a crystallographic plane. 

The multilattice $\mathcal M(\mathbf E; \mathbf w_1, ..., \mathbf w_m)$ can be used for representing the crystal structure as well, i.e. the isometries with the consideration of both the lattice points and atomic/molecular sites. The symmetry of a crystal structure is defined by its \emph{space group}: skeleton Bravais lattice $+$ site symmetry (point group). \cite{hahn1983} In consistency with the symmetry operations defined in the International Tables of Crystallography, the fractional vectors $\mathbf w_i$ should be classified into 1) lattice points; 2) sites. For those Bravais lattices whose primitive bases are consistent with their conventional bases, they only have one lattice point, {\it i.e.} $[0, 0, 0]$ by default. For the rest of Bravais lattices whose primitive bases are different from their conventional bases, they can be expressed mathematically by the multilattice defined in \eqref{eq:mlattice} using the conventional basis vectors and the corresponding fractional atomic position vectors in the conventional unit cell. For examples, the lattice points of the conventional body-centered cubic are $[0, 0, 0]$ and $\frac{1}{2}[1, 1, 1]$; the lattice points of the conventional face-centered cubic are $[0, 0, 0]$, $[\half, \half, 0]$, $[\half, 0, \half]$ and $[0, \half, \half]$.

The meaning of \emph{site} is slightly different from the lattice point. The site of a crystal structure is the spatial position occupied by a real atom/molecule in the unit cell. To distinguish the lattice point from the site, we define a \emph{crystal structure} as:
\begin{equation}\label{eq:xtal}
\mathcal S(\mathbf E, \mathbf w_i; \mathbf s_\alpha) = \{\mathcal M(\mathbf E; \mathbf w_i) + \mathbf E \mathbf s_\alpha: \mathbf s_\alpha \in \mathbb R^3, |\mathbf s_\alpha| \in [0, 1), \alpha = 1, ..., \nu, i = 1, ..., m\}.
\end{equation}
In the above definition, the fractional vectors $\mathbf w_i$, $i = 1, ..., m$ are the lattice points. $\{\mathbf s_1, ..., \mathbf s_\nu\}$ are the sites occupied by atoms in the unit cell framed by the skeleton lattice $\mathcal M(\mathbf E; \mathbf w_i)$. The site symmetry means that the site is invariant under the point group operations including $n-$fold proper rotations, mirrors and reflections. For example, the site $\mathbf s = \epsilon [1, 1, 1]$ where $|\epsilon| < 1$ and its isometries of 
$
\mathcal S(a_0\mathbf I, \mathbf w_1,...,\mathbf w_4; \mathbf s)
$ 
is invariant under $m\bar{3}m$ where $\{\mathbf w_1, ...\mathbf w_4\}$ is given in \eqref{eq:mfcc}. 

In X-ray experiments, the diffraction intensity strongly depends on the scattering of atoms occupying the certain sites of a crystal structure having special site symmetry, although the necessary condition of diffraction ({\it i.e.} Bragg condition) is governed by the underlying periodicity. X-ray analysis software generally requires the knowledge of both the lattice parameters for the skeleton lattice ({\it i.e.} parameters in Table \ref{tb:lp}), and all sites occupied by different species of atoms. In particular, these sites are classified as the \emph{Wyckoff positions} \cite{aroyo2006bilbao} used by crystallographers and most materials scientists. 

For cubic structures, it is not very hard to make an ansatz for the sites in the conventional unit cell. For example, most AB type alloys, the site is among the special positions such as corners, side centers, face centers and body centers. In some cases, atoms of small radii occupy the tetrahedron interstitial sites such as $[\quarter, \quarter, \quarter]$ and $[\quarter, \quarter, \frac{3}{4}]$. However, for some complex crystal structures in low symmetry, {\it e.g.} martensite structures, there is no rational way of obtaining the sites for each of the atoms. From the database of the binary phase diagram \cite{binaryphase2010}, we observe that many low symmetry structures of metallic materials are in fact formed through solid--solid phase transformations from a high temperature phase of cubic symmetry. Examples include steel, CuAl alloy, Nitinol, and many other Cu-based $\beta$ shape memory alloys. Using the mathematical formulation of the crystal structure given in \eqref{eq:xtal}, we can derive the crystal structure of the low symmetry phase from their cubic parent phase through the assigned lattice correspondences.

\subsection{Sublattice and rebase}
When phase transformation occurs, the average lattice distortion can be calculated as a linear transformation that maps the proper lattice basis of parent lattice to that of the product lattice based on the Cauchy-Born rule:
\begin{equation}\label{CB}
\mathbf m_i = \mathbf F \mathbf g_i
\end{equation}
where $\mathbf g_i$ and $\mathbf m_i$ are the lattice vectors in parent and product lattice respectively. Solving these linear equations, we obtain the deformation gradient $\mathbf F = \mathbf{m}_i \otimes \mathbf g^i \in \mathbb R^{3 \times 3}$ where $\mathbf g_i \otimes \mathbf g^i = \mathbf I$.
Chen et al. (2016) \cite{Chen_2016jmps} and Koumatos and Muehlemann (2016, 2017) \cite{Koumatos2016, Koumatos2017} showed that an admissible linear transformation should be consistent with the least transformation strain associated with the proper choice of the correspondent lattice vectors in the parent and product lattices. In crystallography community, the lattice correspondence between phases is understood as the orientation relationship, which usually consists of a set of parallel crystallographic planes and directions in both phases. Here, the transformation strain is calculated as $\sqrt{\mathbf F^T \mathbf F} - \mathbf I$ where $\mathbf F$ is the deformation gradient given by the Cauchy-Born rule in \eqref{CB}.

In real materials, both parent and product phases are the crystal structures mathematically expressed by \eqref{eq:xtal}. 
Suppose the parent phase is a cubic type structure defined as $\mathcal S(a_0\mathbf I, \mathbf w_1, ..., \mathbf w_m; \mathbf s_1, ...,\mathbf s_\nu)$. $m = 1$ for sc, $m = 2$ for bcc, and $m = 4$ for fcc. The skeleton of the product phase can be derived by a linear transformation of a \emph{sublattice} of the parent phase, which underlies a correspondent unit cell of the parent lattice that transforms to the primitive cell of the product lattice. Mathematically, it can be easily proved that a sublattice of Bravais lattice $\mathcal L(\mathbf E)$ is also a Bravais lattice $\mathcal L(\mathbf E \mathbf L)$ where $\mathbf L \in \mathbb Z^{3\times 3}$ and $\det \mathbf L \geq 1$. The three column vectors of $\mathbf L$ are the lattice correspondence vectors for the phase transformation written in the primitive basis $\mathcal L(\mathbf E)$. 
Therefore the point group of $\mathcal L(\mathbf E \mathbf L)$ is a subgroup of the point group of $\mathcal L(\mathbf E)$. From the sublattice $\mathcal L(\mathbf E \mathbf L)$, we can calculate the \emph{reference} lattice parameters for the product phase from its lattice metric tensor:
\begin{equation}\label{eq:derivedmetric}
\Cbar = \mathbf L^T \mathbf C \mathbf L,
\end{equation}
where $\mathbf C$ are the lattice metric tensor of the Bravais lattice of parent phase. Note that the lattice metric $\Cbar$ without any linear perturbation still represents the same symmetry of the parent phase in multilattice setting by putting back all missed lattice points in the frame of the sublattice $\mathcal L(\mathbf E \mathbf L)$ (This will be explained later). 
As an example, we consider a phase transformation from face centered cubic to tetragonal subjected to the lattice correspondence:
\begin{eqnarray}\label{eq:lcv}
&[\frac{1}{2}, \frac{1}{2}, 0]_\text{fcc}& \to [1, 0, 0]_\text{t} \nonumber \\
&[-\frac{1}{2}, \frac{1}{2}, 0]_\text{fcc}& \to [0, 1, 0]_\text{t} \nonumber \\
&[0, 0, 1]_\text{fcc}& \to [0, 0, 1]_\text{t} \nonumber.
\end{eqnarray}
The matrix representation of the lattice correspondence is:
\begin{equation}\label{eq:lc}
\renewcommand*{\arraystretch}{0.75}
\mathbf L = \begin{pmatrix}\frac{1}{2}& -\frac{1}{2}& 0\\\frac{1}{2}& \frac{1}{2}& 0\\0& 0& 1\end{pmatrix}.
\renewcommand*{\arraystretch}{1}
\end{equation}
Note that the components of $\mathbf L$ are integers only when they are written in the primitive lattice basis. The components of \eqref{eq:lc} are in terms of the conventional basis in consistent with the crystallographic expressions mostly used by experimentalists. The component transformation between the primitive and conventional bases follows $[\mathbf L]_p = \chi [\mathbf L]_c$ where $\chi$ is the conversion matrix from the primitive basis of fcc to its conventional basis, defined by \eqref{eq:primitivefcc}. Using \eqref{eq:derivedmetric}, the reference lattice metric for the product lattice before transformation is calculated as:
\begin{equation}\label{eq:lpsq}
\bar{\mathbf C} = [\mathbf L]_p [\mathbf E]_p^T [\mathbf E]_p [\mathbf L]_p = [\mathbf L]_c [\mathbf E]_c^T [\mathbf E]_c [\mathbf L]_c =  \begin{bmatrix}\frac{a_0^2}{2}&0&0\\0&\frac{a_0^2}{2}&0\\0&0&a_0^2\end{bmatrix}.
\end{equation}
Therefore the sublattice parameter sextuples are $(\frac{a_0}{\sqrt{2}}, \frac{a_0}{\sqrt{2}}, a_0, 0, 0, 0)$ before phase transformation. 

For materials undergoing solid--solid phase transformation without chemical doping or reaction, the atoms in the parent crystal structure will neither appear nor disappear upon the phase transformation. But there are lattice points of $\mathcal L(\mathbf E)$ that are not covered by the periodicity of $\mathcal L(\mathbf E \mathbf L)$. 
Let $\mathbf G = \mathbf E \mathbf L$ be the sublattice basis where $\det \mathbf L = m \geq 1$. For any lattice point in $\mathcal L(\mathbf E)$:
\begin{equation}
\mathbf x = \mathbf E \mathbf n = \mathbf G (\mathbf L^{-1} \mathbf n) \text{ for some } \mathbf n \in \mathbb Z^3.
\end{equation}
Since $\det \mathbf L^{-1} = 1/m \leq 1$, the coordinates of the lattice point $\mathbf x$ in basis $\mathbf G$ are not always integers. Those points having fractional coordinates are the missed lattice points under periodicity defined by basis $\mathbf G$. We use the multilattice description to add those missed lattice points back to the unit cell as
\begin{equation}\label{eq:derivedlattice}
\mathcal M = \{\mathbf G (\nbar + \wbar_i): \text{ for all } \nbar \in \mathbb Z^3, |\wbar_i| \in [0, 1),  i = 1, ..., m, \det \mathbf L = m\}.
\end{equation}
The fractional vectors in the frame of the unit cell $\mathcal U(\mathbf G)$ can be calculated by:
\begin{equation}\label{eq:conv}
\wbar_i = \mathbf L^{-1} \mathbf n - \lfloor \mathbf L^{-1} \mathbf n \rfloor, \text{ for all } \mathbf n \in \mathbb Z^3.
\end{equation}
The above calculation is to transform the coordinates of  the missed lattice points in the basis of $\mathbf E$ to the basis of $\mathbf G$, and modulate them back into the unit cell. We define this operation as \emph{rebase}.
Using the primitive lattice basis of fcc lattice and the transformation matrix $[\mathbf L]_p$ with $\det [\mathbf L]_p = 2$, by equation \eqref{eq:conv}, the multilattice after rebasing consists of two lattice points $[0, 0, 0]$ and $[\frac{1}{2}, \frac{1}{2}, \frac{1}{2}]$. The derived lattice looks like a body centered tetragonal with lattice parameters $(\frac{a_0}{\sqrt{2}}, \frac{a_0}{\sqrt{2}}, a_0, 0, 0, 0)$, it still presents the fcc symmetry as discussed earlier. To break and obtain a lower symmetry, we need to perturb the periodicity of $\mathcal M(\mathbf G; [0,0,0], [\frac{1}{2}, \frac{1}{2}, \frac{1}{2}])$ by a linear transformation:
\begin{eqnarray}\label{eq:P}
\mathbf P &=& \frac{\delta_1}{\mathbf g_1 \cdot \mathbf g_1} \mathbf g_1 \otimes \mathbf g_1 + \frac{\delta_2}{\mathbf g_2 \cdot \mathbf g_2} \mathbf g_2 \otimes \mathbf g_2 + \frac{\delta_3}{\mathbf g_3 \cdot \mathbf g_3} \mathbf g_3 \otimes \mathbf g_3\\
& + & \frac{\gamma_{12}}{\sqrt{(\mathbf g_1 \cdot \mathbf g_1)(\mathbf g_2 \cdot \mathbf g_2)}} \mathbf g_1 \otimes \mathbf g_2 + \frac{\gamma_{13}}{\sqrt{(\mathbf g_1 \cdot \mathbf g_1)(\mathbf g_3 \cdot \mathbf g_3)}} \mathbf g_1 \otimes \mathbf g_3 + \frac{\gamma_{23}}{\sqrt{(\mathbf g_2 \cdot \mathbf g_2)(\mathbf g_3 \cdot \mathbf g_3)}} \mathbf g_2 \otimes \mathbf g_3,
\end{eqnarray}
in which the three vectors $(\mathbf g_1, \mathbf g_2, \mathbf g_3) = \mathbf G$, the parameters $\delta_i, i = 1, 2, 3$ are the self-assigned stretches and the parameters $\gamma_{12}$, $\gamma_{13}$ and $\gamma_{23}$ are the self-assigned shears. Physically, these stretch and shear quantities are small values, which underlie the deformation of the reference lattice metric of the parent phase to the metric of the product phase.
 
For complex parent crystal structures such as alloys and compounds, we can use the \emph{substructure} $\mathcal S(\mathbf P \mathbf G, \wbar_i; \sbar_\alpha)$ as the derived crystal structure of the product phase. Both lattice points and sites are rebased and transformed by the same calculations using \eqref{eq:conv} and \eqref{eq:P}. The  derived lattice parameters and the atomic positions in the unit cell can be used to define the unknown structure of low symmetry from solid-solid phase transformation, which provides the initial conditions for structural refinement in X-ray diffraction analysis.  

\section{Orientation relationship and parallelism}
In this section, we work on the Bravais lattice index only.\footnote{Here, the \emph{index} is equivalent to the coordinate of a lattice vector.} That is the integer tuple $\mathbf n = (n_1, n_2, n_3) \in \mathbb Z^3$ of a Bravais lattice $\mathcal L(\mathbf E_p)$ where $\mathbf E_p$ is the primitive lattice basis. Therefore, it is not always referred to the Miller index of crystallographic direction, especially for the lattices whose primitive lattice bases are not consistent with the conventional bases such as bcc, fcc, bct, bco, fco, ico and bcm (see Table \ref{tb:lp}). Let $\mathbf E_c$ denote the conventional lattice basis. The primitive lattice basis is related to the conventional basis by $\mathbf E_p = \mathbf E_c \chi$ where $\chi \in \mathbb R^{3 \times 3}$ is the conversion matrix between two bases.
For fcc lattice,
\begin{equation}\label{eq:chifcc}
    \chi_\text{fcc} = \begin{bmatrix}\half & 0 & \half\\\half & \half & 0\\0 & \half & \half\end{bmatrix}.
\end{equation}
For bcc lattice,
\begin{equation}\label{eq:chibcc}
    \chi_\text{bcc} = \begin{bmatrix}\half & \half & -\half\\-\half & \half & \half\\\half & -\half & \half\end{bmatrix}.
\end{equation}
The presentation of lattice basis varies upon the symmetry and/or lattice invariant transformations, therefore the expressional examples given in \eqref{eq:chifcc} and \eqref{eq:chibcc} may vary as the change of lattice basis.
Using the conversion matrix, the Bravais lattice index is related to the Miller index ($\mathbf n_c$) of crystallography by $\mathbf n = \chi^{-1} \mathbf n_c$. Under the rebase transformation by lattice correspondence matrix $[\mathbf L]_p$, the index of the lattice vector in the basis of the sublattice $\mathcal L(\mathbf E_p [\mathbf L]_p)$ is calculated as
\begin{equation} \label{eq:sub-conv}
\mathbf n_\text{s} = [\mathbf L]_p^{-1} \mathbf n = (\chi [\mathbf L]_p)^{-1} \mathbf n_c = ([\mathbf L]_c)^{-1} \mathbf n_c.
\end{equation} 
The equation \eqref{eq:sub-conv} underlies the transformation rule for the Miller index of crystallography from the Bravais lattice to its derived lattice. Through solid-solid phase transformation, the skeleton periodicity of the derived lattice changes for the energetic reasons, but the index of the product lattice remains the same relationship with the parent lattice. Due to the small lattice distortion, instead of the equal sign suggested in \eqref{eq:sub-conv}, such a lattice index relationship becomes the parallelism 
\begin{equation}\label{eq:transvec}
\mathbf n_\text{dv} \ ||\ ([\mathbf L]_c)^{-1} \mathbf n_c.
\end{equation} 
For the crystallographic plane, the lattice plane index relationship can be expressed as the parallelism
\begin{equation}\label{eq:transplane}
\mathbf n^\ast_\text{dv}\ ||\ ([\mathbf L]_c)^T \mathbf n^\ast_c,
\end{equation}
in which, $\mathbf n^\ast_\text{dv}$ and $\mathbf n^\ast_c$ are the Miller indicies of the crystallographic planes for the derived and parent lattices respectively. 

By \eqref{eq:transvec}, \eqref{eq:transplane} and the Cauchy-Born rule in \eqref{CB}, we generalize the presentation of orientation relationship and the associated transformation strains for solid-solid phase transformation through a unified quantity: the lattice correspondence matrix. This correspondence matrix can be rigorously determined by the StrucTrans algorithm \cite{Chen_2016jmps}, which minimizes the transformation strains between two lattices of arbitrary symmetries.

\section{Structural determination of martensite CuAlMn}

Among all shape memory alloys, CuAuZn, CuAlNi and CuAlZn alloy systems form the second largest group of material candidates used in research and development. However, their commercialization is highly confined due to the poor fatigue life when they are in a polycrystalline state. 
Compared to these Cu-based alloy systems, the CuAlMn system is much less developed. For some aluminum compositions, people demonstrated good ductility in this alloy system \cite{sutou2008ductile}. In-depth study of the crystallography and the formation of microstructure is highly hindered in this system because of the lack of the structural parameters for martensite. Some discussions about its micromechanical behaviors are based on the lattice parameters of the cubic to monoclinic transformation of its sibling system CuAlNi \cite{wang2002ebsd, fornell2017orientation}. In this section, we will show how the derived lattice assists the analysis of the structural determination of martensite structure of CuAlMn alloy. 

\begin{figure}[ht]
\centering
\includegraphics[width=3.6in]{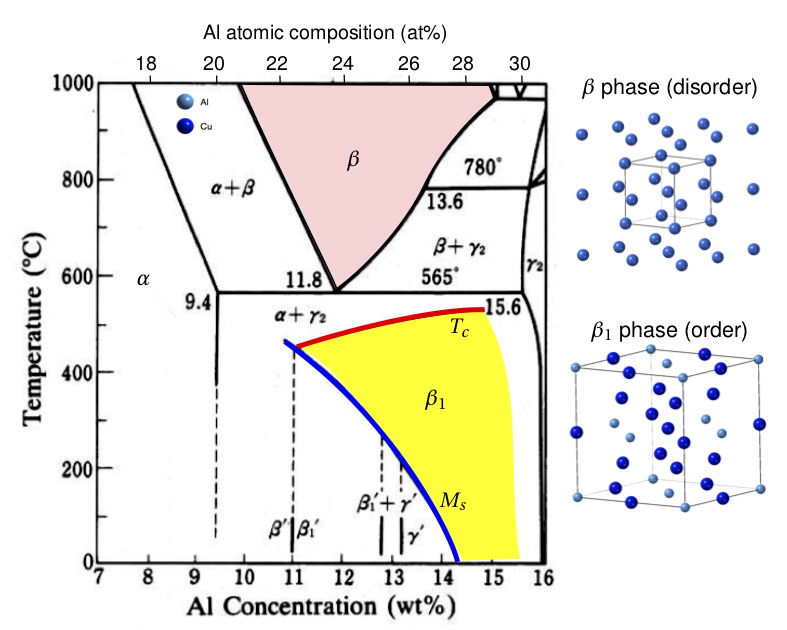}
\caption{Binary phase diagram of Cu-Al alloy from FScopp alloy database 2012. }\label{fig:cu-phase}
\end{figure}

According to the binary phase diagram of Cu-Al alloy system, the martensite can be induced by suppressing the eutectic transformation through the rapid cooling process. The high temperature $\beta$ phase directly transforms into an ordered structure $\beta_1$ (DO3, or L2$_1$) at $T_c$ (marked as the red curve in Figure \ref{fig:cu-phase}). The $\beta_1$ phase further undergoes a martensitic transformation at temperature $M_s$ marked as the blue curve in Figure \ref{fig:cu-phase}. This also works for ternary Cu-based $\beta$ alloys doped by Mn \cite{sutou1999}, except that the $\beta_1$ zone (the yellow region in Figure \ref{fig:cu-phase}) and the composition dependent $M_s$ curve may vary with the addition of Mn. The symmetry of the product phase formed by martensitic transformation from the $\beta_1$ phase varies with the Al concentration \cite{warlimont1974, sutou1999, sutou2002lattparam}.

The most studied compositions of Cu-Al-Mn are those with Al compositions 14 - 17 at\% and Mn composition around 10 at\% \cite{kainuma1996, sutou1999, sutou2002lattparam, sutou2004} since the alloys within this compositional range show a better ductility than CuAlNi and CuAlZn in polycrystalline form \cite{sutou1999, sutou2008ductile}. The crystal structure of martensite has been characterized by both X-ray powder diffraction and transmission electron microscopy for Al compositions 14, 16 and 17 at\%. It was found that the martensite of as-aged samples are 18R ({\it i.e.} 18 layers modulated monoclinic structure).  In our study, we choose the alloy Cu$_{67}$Al$_{24}$Mn$_9$ (at\%) to demonstrate our derived lattice theory, since there is a discrepancy in the understanding of martensitic structure in this composition domain\cite{Obrado_1997}. 

\subsection{Experiment}
A mixture of high purity Cu (99.99 wt\%), Al (99.999 wt\%), and Mn (99.95 wt\%) ingots were melted in a quartz tube placed in an evacuated ($10^{-5}$ mbar) induction furnace under argon atmosphere. The melt was injected into a cylindrical copper mold and solidified as a rod of diameter 5 mm. It was homogenized at 800$^\circ$C for 3 hours under argon atmosphere, then cooled down in the furnace. We cut the rod into thin slices of thickness 1 mm, which were sealed in a vacuum quartz tube, heat treated at 900$^\circ$C for 1 hour and quenched in water. 

\begin{figure}[ht]
\centering
\includegraphics[width=3.in]{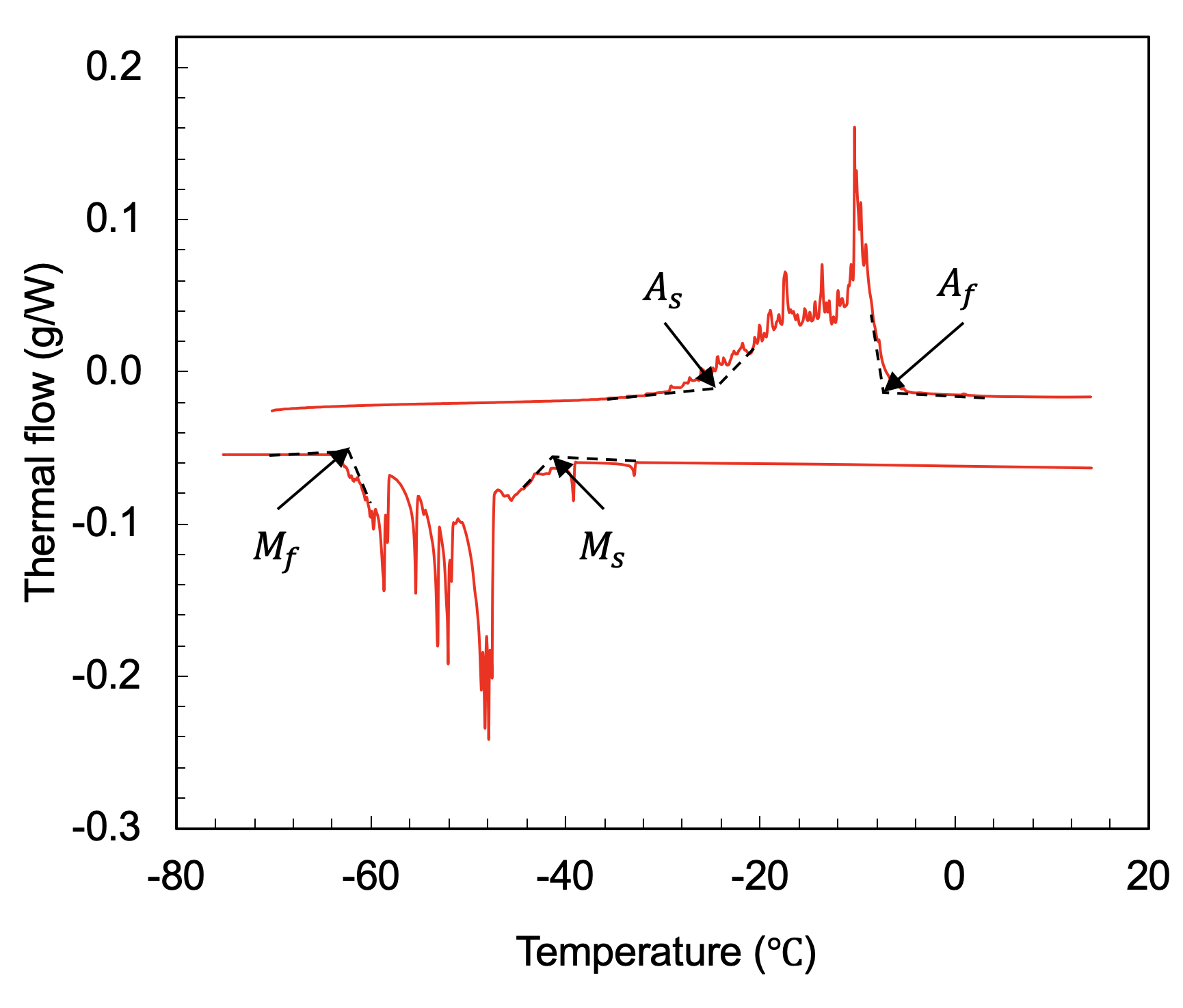}
\caption{Differential Scanning Calorimetry of CuAl$_{24}$Mn$_9$. }\label{fig:dsc}
\end{figure}

The transformation temperature of the specimen was measured by differential scanning calorimetry (DSC) by TA Instruments Q1000 at a heating and cooling rate of 10$^\circ$C/min for three complete thermal cycles in the range from -75$^\circ$C to 20$^\circ$C. The austenite start/finish $A_s$/$A_f$ temperatures and martensite start/finish $M_s$/$M_f$ temperatures are determined as the onsets of the heat absorption/emission peaks as shown in Figure \ref{fig:dsc}: $A_s = -26^\circ$C, $A_f = -8^\circ$C, $M_s = -48^\circ$C and $M_f = -62^\circ$C. The thermal hysteresis is quite large and measured to be $\Delta T = \frac{1}{2}(A_s + A_f - M_s - M_f) = 38^\circ$C. Unlike those reported transforming Cu-based $\beta$ alloys \cite{kainuma1996, Mallik_2008}, this one shows the thermal bursts in large magnitude over a wide temperature range during the phase transformation. Evidently this thermal signature was observed in a close-by alloy system with slightly different mangenese compositions \cite{Obrado_1997}, but the detailed crystal structure of martensite has not been thoroughly studied for this series. In their work, the total entropy change from the cubic phase (austenite) is calculated based on the DSC measurements, which was found to be highly correlated to the average electron concentration ({\it i. e.} $e/a$) of the alloy. Only those with $e/a > 1.46$ showed the jerky thermal behaviors, which was conjectured as a different type of martensitic transformation: bcc to 2H. Their room-temperature powder diffraction measurement showed some residual peaks corresponding to the 2H structure. The martensite finish temperatures of this series are quite low, {\it i.e.} around $-60^\circ$C, therefore the room-temperature diffractometry with mixed phases is not sufficient to identify and solve the martensite crystal structure for the 2H phase, nor the report of lattice parameters were reported for this phase in any other CuAlMn system. 

%

\subsection{Advanced structural characterization by synchrotron X-ray microdiffraction}
To obtain precise information for the crystal structures of austenite and martensite and to show
how the derived lattice theory assists the structure determination, we conducted a temperature-varying single crystal synchrotron X-ray Laue microdiffraction  experiment combined with monochromatic energy scans \cite{Tamura_2014book} at beamline 12.3.2 of the Advanced Light Source, Lawrence Berkeley National Lab. The X-ray beam with energy bandpass from 6 keV to 24 keV was focused down to 1 $\mu$m in diameter by a pair of elliptically bent Kirkpatrick-Baez mirrors. The focused high-brightness X-rays illuminated a single grain of the bulk sample, and generated a single crystal Laue pattern. We used the custom-made thermal stage \cite{Chen_2016APL} to drive the phase transformation of the bulk sample, which controls the sample temperature from -100$^\circ$C to 200$^\circ$C with ramping rate of 15$^\circ$C/min. The bulk sample was polished in the austenite form at room temperature. An optical microscope attached to the end-station optic box allows to observe in-situ the sample surface reliefs while collecting the Laue patterns at a specified sample position during cooling process. 

\begin{figure}[h]
\centering
\includegraphics[width=0.65\textwidth]{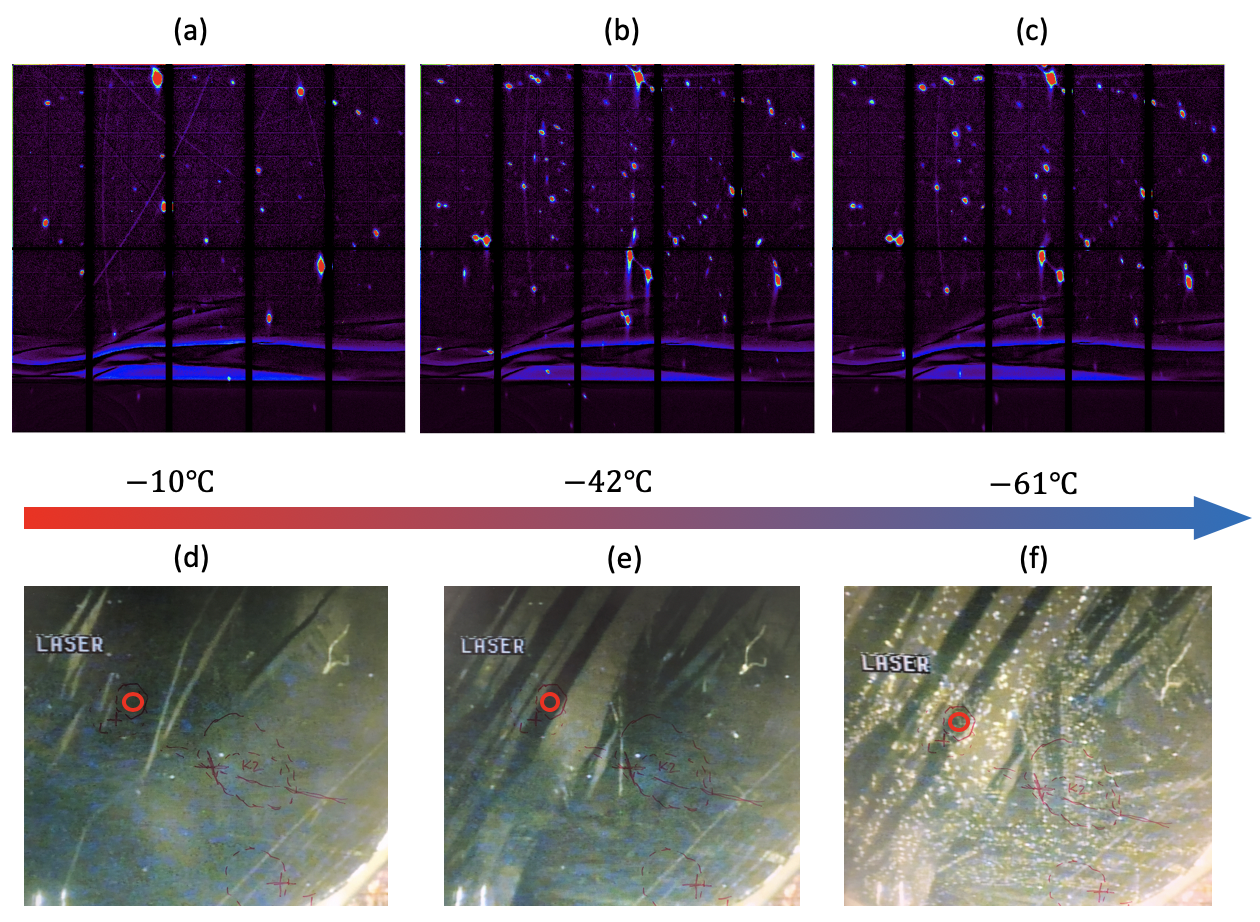}
\caption{(a) -- (c) Laue diffraction patterns of the bulk sample from high temperature to low temperature corresponding to the microstructures in (d) -- (f), in which the red circles denote the surface positions illuminated by the focused X-ray beam. } 
\label{fig:laue}
\end{figure}

Figures \ref{fig:laue} (a) -- (c) show the evolution of Laue patterns as the sample was cooled down through the phase transformation temperature while the corresponding microstructures in Figure \ref{fig:laue} (d) -- (f) sufficiently reveal that the Laue pattern in (a) purely represents the austenite phase. As the temperature going down, we observed that martensite laths appear and grow as shown in Figure \ref{fig:laue} (e) and (f). The Laue patterns (b) and (c) suggest that they are purely in the martensitic phase. 

We used the L2$_1$ structure (space group F$m\bar{3}m$) to index the austenite Laue pattern. The crystal structure is depicted in Figure \ref{fig:austenite}(b). \cite{Tilley_book} The stoichiometric ratio of atoms for L2$_1$ is supposed to be ABC$_2$ where A atoms occupy the site 4a at $[0, 0, 0]$ with site symmetry $m\bar{3}m$, B atoms occupy the site 4b at $[\frac{1}{2}, \frac{1}{2}, \frac{1}{2}]$ also with site symmetry $m\bar{3}m$, and C atoms occupy 8c site at $[\frac{1}{4}, \frac{1}{4}, \frac{1}{4}]$ with site symmetry $\bar43m$. In the case of Cu$_{67}$Al$_{24}$Mn$_9$, we assume that the Al atoms fully occupy the 4a site,  the Mn atoms fully occupy the 4b site and the Cu atoms occupy the 8c sites. Using the XMAS software \cite{Tamura_2014book}, we successfully indexed the Laue pattern by the proposed L2$_1$ structure as shown in Figure \ref{fig:austenite}(a). To determine the austenite lattice parameter, we chose four $(hkl)$ reflections: $(252)$, $(170)$, $(238)$ and $(176)$, and precisely measure their interplanar distances by performing energy scans of the reflections. The refined lattice parameter was measured to be $a_0 = 5.87897 \AA$. 

\subsection{Determination of martensitic structure by derived lattice theory}
We assume the Bain lattice correspondence for the phase transformation from L2$_1$ to orthorhombic given in :
\begin{equation}\label{eq:correspondence}
\mathbf L = \begin{bmatrix}\frac{1}{2}&0&-\frac{1}{2}\\0&1&0\\\frac{1}{2}&0&\frac{1}{2}\end{bmatrix}
\end{equation}
and compute the sublattice points of $\mathcal L(a_0 \mathbf L)$ for the 4a site $[0, 0, 0]$, 4b site $[\frac{1}{2}, \frac{1}{2}, \frac{1}{2}]$ and 8c site $[\frac{1}{4}, \frac{1}{4}, \frac{1}{4}]$ respectively using the rebase equation \eqref{eq:conv}.

\begin{figure}[h]
\centering
\includegraphics[width=0.65\textwidth]{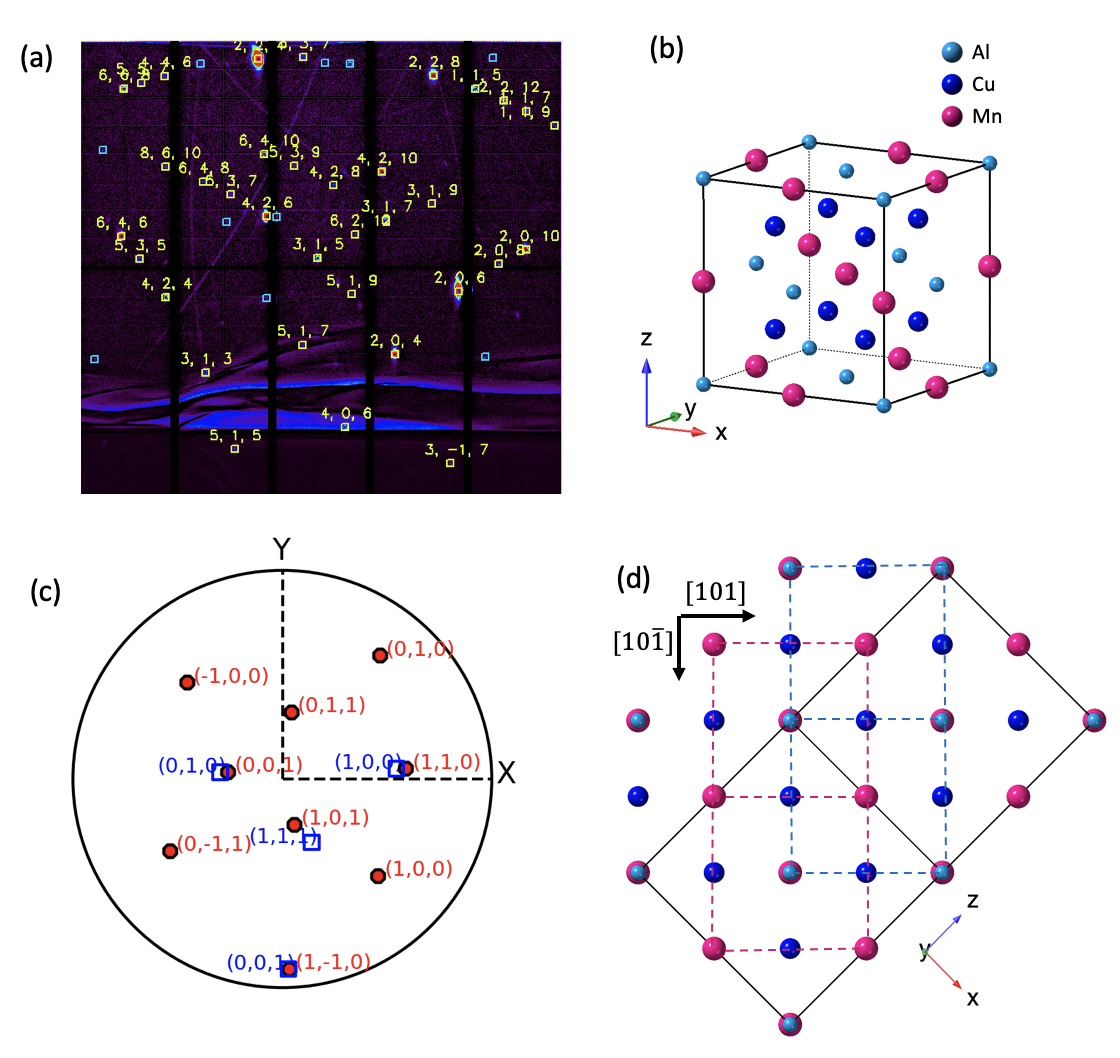}
\caption{Austenite crystal structure of Cu$_{67}$Al$_{24}$Mn$_9$: (a) Indexed Laue pattern by (b) the L2$_1$ structure. (c) The stereographic projection of the reciprocal lattices of austenite and the transformed martensite with respect to the stage coordinate system (X -- Y -- Z). (d) The theoretical atomic structure in $(010)$ plane that is aligned horizontally along $[101]$ direction. } 
\label{fig:austenite}
\end{figure}

\begin{figure}[h]
\centering
\includegraphics[width=0.65\textwidth]{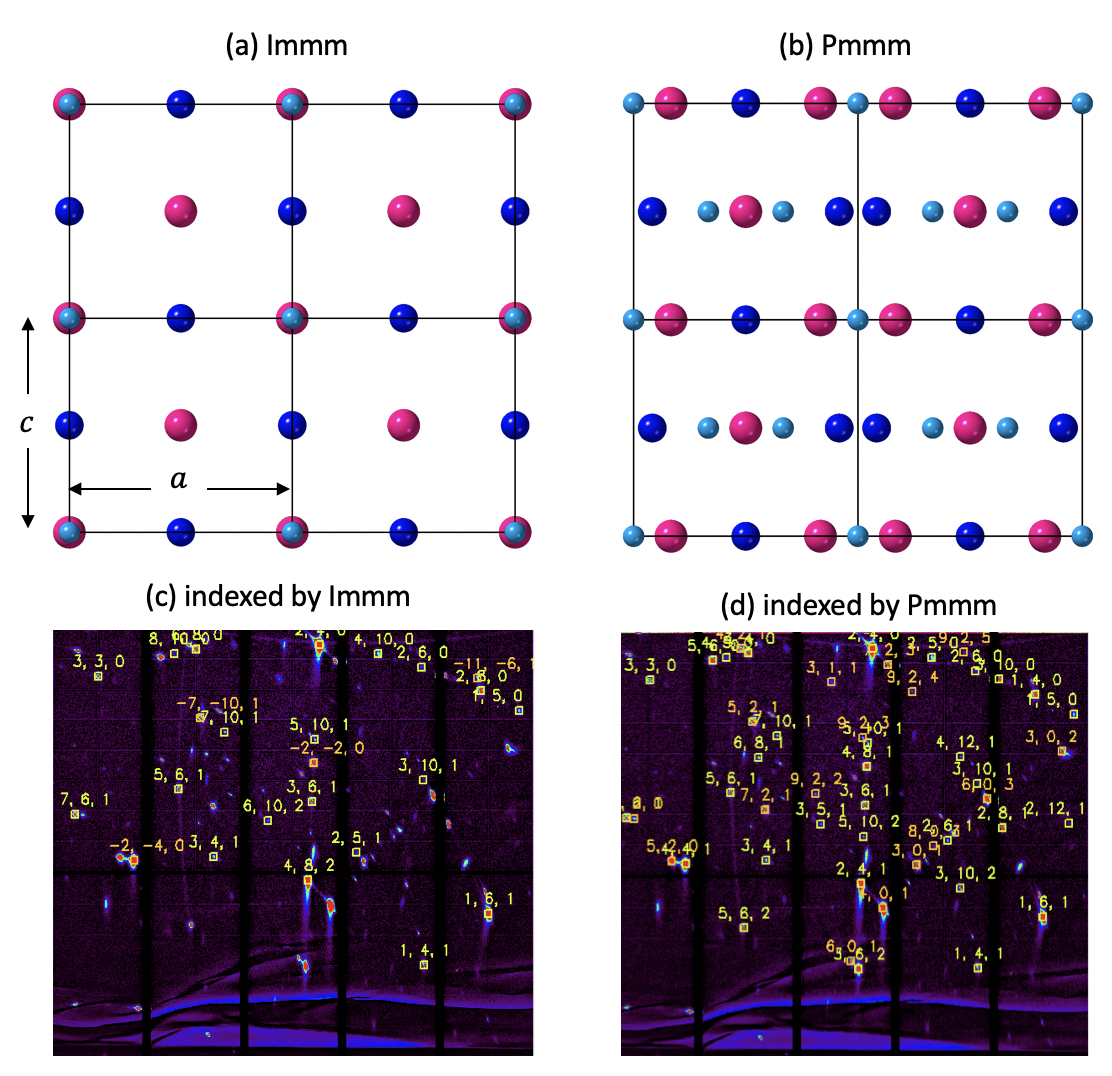}
\caption{Derived orthorhombic structures of CuAl$_{24}$Mn$_9$ with space group (a) I$mmm$ and (b) P$mmm$, by which the synchrotron X-ray Laue diffraction pattern of martensite is indexed in (c) and (d) respectively.} 
\label{fig:martensite}
\end{figure}

After rebasing the original L2$_1$ lattice, the sublattice structure $\mathcal S = \{\mathcal L(a_0 \mathbf L) + a_0 \mathbf L \mathbf s_\alpha: \text{ for } \alpha = 1, ...,8\}$ retains its stoichiometry with sites for :
\begin{eqnarray}
&\text{Al: }& \mathbf s_1 = [0, 0, 0], \ \mathbf s_2 = [\frac{1}{2}, \frac{1}{2}, \frac{1}{2}],\nonumber\\
&\text{Mn: }& \mathbf s_3 = [\frac{1}{2}, 0, \frac{1}{2}], \ \mathbf s_4 = [0, \frac{1}{2}, 0],  \label{eq:atompos}\\
&\text{Cu: }& \mathbf s_5 = [\frac{1}{2}, \frac{1}{4}, 0], \ \mathbf s_6 = [0, \frac{3}{4}, \frac{1}{2}], \ \mathbf s_7 = [\frac{1}{2}, \frac{3}{4}, 0], \ \mathbf s_8 = [0, \frac{1}{4}, \frac{1}{2}].  \nonumber
\end{eqnarray}
By equation \eqref{eq:lpsq}, the sublattice parameters are $(\bar a, \bar b, \bar c, 0, 0, 0)$ where $\bar a = \bar c = \frac{a_0}{\sqrt{2}} = 4.15706 \AA$ and $\bar b = a_0 = 5.87897 \AA$. In Figure \ref{fig:austenite} (d), the red and blue boxes underlie the sublattice cells of $\mathcal L(a_0\mathbf L)$.  To generate the derived orthorhombic structure, we propose a simple stretch tensor :
\begin{equation}
\mathbf P = \sum_i \frac{\delta_i}{\mathbf p_i \cdot \mathbf p_i} \mathbf p_i \otimes \mathbf p_i,
\end{equation}
with the three principle stretches $\delta_1 = 1.05$, $\delta_2 = 1.02$ and $\delta_3 = 0.92$ along the principle axes $\mathbf p_i$, which are aligned with the basis of the sublattice $\mathcal L(a_0\mathbf L)$. Then the lattice parameters of the derived substructure are $(a_\text{dv}, b_\text{dv}, c_\text{dv}, 0, 0, 0)$ where $a_\text{dv} = 4.36491 \AA$, $b_\text{dv} = 5.40865 \AA$ and $c_\text{dv} = 4.2402 \AA$. By observation, the sites given in \eqref{eq:atompos} imply a body centered symmetry, which can be naturally expressed as the space group I$mmm$ (number in international table: 71). The derived substructure can be fully described by the Wyckoff positions listed in Table \ref{tab:Immm} \footnote{Wyckoff position is equivalent to the definition of site in this paper.}. Multiplicity of the Wyckoff position means the number of equivalent sites under the site symmetry operations. Using our definition of crystal structure, this derived I$mmm$ structure plotted in Figure \ref{fig:martensite}(a) can be expressed as :
\begin{equation}\label{eq:structImmm}
\mathcal S_\text{I$mmm$} = \{\mathcal M(\mathbf E_\text{orth}; \mathbf w_1, \mathbf w_2) + \mathbf E_\text{orth}\mathbf s_\alpha: \alpha = 1, 2, 3 \}
\end{equation}
where the structural parameters are given by :
\begin{equation}\label{eq:lp_orth}
\mathbf E_\text{orth} = \begin{bmatrix}a_\text{dv}&0&0\\0&b_\text{dv}&0\\0&0&c_\text{dv}\end{bmatrix}, \mathbf w_1 = \begin{bmatrix}0\\0\\0\end{bmatrix}, \mathbf w_2 = \begin{bmatrix}\half\\\half\\\half\end{bmatrix}, 
\end{equation}
with atom sites $\mathbf s_1 = \mathbf w_1$ for Al, $\mathbf s_2 = [\half, 0, \half]$ for Mn and $\mathbf s_3 = [0, \quarter, \half]$ for Cu. 

\begin{table}[h]
\centering
\caption{Wyckoff positions of derived I$mmm$ structure \cite{cryst_wyckoff} for CuAlMn}
\label{tab:Immm}
\begin{tabular}{c|c|c|c}
Multiplicity & Wyckoff letter & Site symmetry & Site with lattice points: $[0,0,0] + [\frac{1}{2}, \frac{1}{2}, \frac{1}{2}]$ \\\hline
2 & a & $mmm$ & $[0, 0, 0]$\\ \hline
2 & d & $mmm$ & $[\frac{1}{2}, 0, \frac{1}{2}]$ \\\hline
4 & h & $m2m$ & $[0, \frac{1}{4}, \frac{1}{2}]$\\\hline
\end{tabular}
\end{table}

We use this crystal structure as input for the XMAS Laue indexing algorithm, and get the indexed Laue pattern of martensite. The indexing program suggests two martensite variants corresponding to the indexed reflections marked by yellow and orange colors respectively  in Figure \ref{fig:martensite}(c). However, the indices of many major reflections are still not found by the crystal structure $\mathcal S_\text{I$mmm$}$, which implies the existence of an even lower symmetry structure. To slightly lower the symmetry of $\mathcal S_\text{I$mmm$}$, we propose to introduce shuffles as indicated in many Cu-based alloys \cite{warlimont1974}. Here we propose the following shear tensor :
\[
\mathbf V = \mathbf I + \frac{-1}{3} [1, 0, 0]^T \otimes [0, 1, 0]^T
\] 
that directly shuffles the sites of structure $\mathcal S_\text{I$mmm$}$ to :
\begin{eqnarray}
&\text{Al: }& \sbar_1 = [0,0,0], \sbar_2 = [\frac{1}{3}, \half, \half] \nonumber \\
&\text{Mn: }& \sbar_3 = [\half, 0, \half], \sbar_4 = [\frac{5}{6}, \half, 0] \label{eq:shuffles} \\
&\text{Cu: }& \sbar_5 = [\half, \quarter, 0], \sbar_6 = [\frac{5}{6}, \quarter, \half]. \nonumber
\end{eqnarray}
The above atom sites do not show the body centered symmetry any more, therefore the underlying periodicity reduces to a primitive orthorhombic symmetry. In the international table of crystallography, the highest symmetry corresponding to the primitive orthorhombic symmetry is P$mmm$ (number: 47), of which the crystal structure is shown in Figure \ref{fig:martensite}(b), expressed as :
\begin{equation}\label{eq:struct_pmmm}
\mathbf S_{\text{P}mmm} = \{\mathcal L(\mathbf E_\text{orth}) + \mathbf E_\text{orth}\sbar_\alpha: \alpha = 1,..., 6 \}
\end{equation}

\begin{table}[h]
\centering
\caption{Wyckoff positions of derived P$mmm$ structure \cite{cryst_wyckoff} for CuAlMn}
\label{tab:pmmm}
\begin{tabular}{c|c|c|c}
Multiplicity & Wyckoff letter & Site symmetry & Site with lattice points: $[0,0,0]$ \\\hline
1 & a & $mmm$ & $[0, 0, 0]$\\ \hline
2 & l & $2mm$ & $[\frac{1}{3}, \half, \half]$ \\\hline
1 & d & $mmm$ & $[\half, 0, \half]$\\\hline
2 & k & $2mmm$ & $[\frac{5}{6}, \half, 0]$\\\hline
2 & o & $m2m$ & $[\half, \quarter, 0]$\\\hline
4 & z & ..$m$ & $[\frac{5}{6}, \frac{1}{4}, \frac{1}{2}]$\\\hline
\end{tabular}
\end{table}

The site symmetry and corresponding Wyckoff positions are listed in Table \ref{tab:pmmm}, with which the chemical stoichiometry of Al, Mn and Cu still maintains $1:1:2$ in consistent with austenite. Using the crystal structure $\mathcal S_{\text{P}mmm}$ as the input for XMAS Laue indexing algorithm, we get the indexed Laue pattern of martensite in Figure \ref{fig:martensite}(d). All major reflections are indexed by two martensite variants, which implies the possible crystal structure of martensite is likely to be $\mathcal S_{\text{P}mmm}$ given by \eqref{eq:struct_pmmm}. 

Finally, the orientation relationship between the austenite (F$m\bar{3}m$) and the derived martensite (P$mmm$) is confirmed by overlapping their stereographic projections calculated from the indexed Laue patterns, shown in Figure \ref{fig:austenite} (c). The normal vectors of the crystallographic planes $(001)$, $(1\bar10)$ and $(110)$ in austenite lattice are parallel to those of the crystallographic planes $(010)$, $(001)$ and $(100)$ in the lattice of one of the martensite variants. This result also confirms our conjecture of the lattice correspondence for the derived martensite structure by \eqref{eq:correspondence}. 

\begin{table}[h]
\centering
	\caption{The results of monochromatic energy scan and the corresponding indices obtained from numerical analysis for martensite phase.} 
\renewcommand{\arraystretch}{1}	
	\begin{tabular}{c c c c c c c}
	\hline
	$hkl$ & $E$ (kev) & $\lambda$ (\textup{\AA}) & $2\theta\;(^{\circ})$ & $d_{exp}$ (\textup{\AA})& $d_{theo}$ (\textup{\AA}) & $d_{exp}/d_{theo}$\\
	\hline
   ($5\bar{4}\bar{1}$) & 11.0201 & 1.12507 & 101.1893 & 0.728038 & 0.727824 & 1.00029\\ 
   ($6\bar{4}\bar{2}$) & 14.4006 & 0.860965 & 88.1896 & 0.618645 & 0.618648 & 1.00000\\
   ($5\bar{4}\bar{2}$) & 13.4255 & 0.923498 & 82.8361 & 0.697982 & 0.697949 & 1.00005\\
   ($4\bar{3}\bar{2}$) & 12.0003 & 1.033176 & 73.7746 & 0.860631 & 0.860778 & 0.99983\\
   ($3\bar{2}\bar{2}$) & 11.3352 & 1.093798 & 59.2741 & 1.105955 & 1.105480 & 1.00043\\
   ($5\bar{2}\bar{2}$) & 12.6406 & 0.980841 & 77.6421 & 0.782307 & 0.782578 & 0.99965\\
   ($7\bar{2}\bar{1}$) & 13.1555 & 0.942451 & 101.2271 & 0.609698 & 0.609752 & 0.99991\\
   ($5\bar{2}\bar{1}$) & 9.9452 & 1.246674 & 98.0783 & 0.825439 & 0.825409 & 1.00004\\
   ($4\bar{1}\bar{2}$) & 12.0503 & 1.028889 & 64.3184 & 0.966504 & 0.966894 & 0.99960\\    
   $(60\bar1)$ & 12.3899 & 1.000688 & 86.8573 & 0.727825 & 0.727816 & 1.00001\\
   $(50\bar1)$ & 10.7153 & 1.157076 & 83.5984 & 0.867996 & 0.867833 & 1.00019\\
   $(40\bar1)$ & 9.1151 & 1.360207 & 78.7222 & 1.072370 & 1.072365 & 1.00000\\   
	\hline
	\end{tabular}
	\label{tab:lattparam}
	\end{table}
	
In the pure martensite phase, we conduct the monochromatic energy scan in a wide photon energy spectrum ({\it i.e.} from 8keV to 16keV), and get a list of interplanar distance measures in Table \ref{tab:lattparam} for the indexed crystallographic planes in a reference Laue pattern. In general, the theoretical interplanar distance for a plane $\mathbf h = (hkl)$ can be expressed as $d_{theo} = |\mathbf E^\ast \mathbf h|$ where $\mathbf E^\ast = \mathbf E^{-T}$. Here $\mathbf E^\ast$ is the reciprocal lattice basis for the Bravais lattice $\mathcal L(\mathbf E)$. In the case of orthorhombic lattice, the lattice basis $\mathbf E$ is a diagonal matrix with diagonal elements $(a, b, c)$. The values of $(a, b, c)$ are determined as the minimizers 
\begin{equation}
    (a^\ast, b^\ast, c^\ast) = \text{arg}\min_{(a, b, c) \in \mathbb R^3} \sum_{\mathbf h \in \mathcal H}\Vert d_{theo}(a, b, c; \mathbf h) - d_{exp}(\mathbf h) \Vert^2,
\end{equation}
in which the set $\mathcal H$ consists of all selected $\mathbf h = (hkl)$ corresponding to the experimentally measured $d_{exp}(\mathbf h)$ listed in Table \ref{tab:lattparam}.
We use the derived lattice parameters $(a_\text{dv}, b_\text{dv}, c_\text{dv})$ as the initial condition and get the refined lattice parameters $a=4.43196\AA$, $b=5.34533\AA$, $c=4.26307\AA$. Using the refined lattice parameters, we calculated the interplanar distances for the $(hkl)$ planes shown in Table \ref{tab:lattparam}, which agree with the measured values up to $0.01\%$.

\section{Conclusion}
Using the mathematical description of the crystal structure: Lattice + atom sites in consistent with the international table of crystallography, we underlie a route to calculate the derived crystal structure for the martensitic materials transformed from solid-solid phase transformation. This approach is useful for the structure determination by X-ray diffraction analysis. Not only it provides the heuristic lattice parameters by the small perturbation based on continuum mechanics theory, but it also calculates the atom positions and their site symmetries based on the Cauchy-Born rule from the orientation relationship. We use a typical Cu-based $\beta$ alloy to demonstrate our theory and showed that the indexing accuracy of X-ray Laue pattern has been improved by a large margin based on our mathematical framework. The derived lattice theory can be used in general X-ray analysis when the unknown material is in low symmetry phase, which has some inherit signature with the high symmetry phase. Finally, we show that the precise lattice parameters of the 2H martensite of Cu$_{67}$Al$_{24}$Mn$_{9}$ alloy can be determined without pre-knowledge of the structure.

\section{Acknowledgement}
M. K and X.C. thank the HK Research Grants Council for financial support under Grants No. 26200316 and No. 16207017. X.C. also thanks the Isaac Newton Institute for Mathematical Sciences for support and hospitality during the program “The Mathematical Design of New Materials,” when work on this paper was undertaken. This work was supported by EPSRC Grant No. EP/R014604/1. The research of YY is supported by City University of Hong Kong with the grant number 9610391. Beamline 12.3.2 and the Advanced Light Source were supported by the Office of Science, Office of Basic Energy Sciences, of the U.S. Department of Energy under Contract no. DE-AC02-05CH11231.

\pagebreak
\bibliography{struct_19_ref}

\begin{thebibliography}{10}
\expandafter\ifx\csname url\endcsname\relax
  \def\url#1{\texttt{#1}}\fi
\expandafter\ifx\csname urlprefix\endcsname\relax\def\urlprefix{URL }\fi
\expandafter\ifx\csname href\endcsname\relax
  \def\href#1#2{#2} \def\path#1{#1}\fi

\bibitem{chang1951}
L.~Chang, T.~Read, {Behavior of the elastic properties of AuCd}, Trans Met Soc
  AIME 191 (1951) 47.

\bibitem{tadaki1998cu}
T.~Tadaki, {Cu-based shape memory alloys}, Shape memory materials (1998)
  97--116.

\bibitem{miyazaki1982characteristics}
S.~Miyazaki, Y.~Ohmi, K.~Otsuka, Y.~Suzuki, {Characteristics of deformation and
  transformation pseudoelasticity in Ti-Ni alloys}, Le Journal de Physique
  Colloques 43~(C4) (1982) C4--255.

\bibitem{otsuka2005physical}
K.~Otsuka, X.~Ren, {Physical metallurgy of Ti--Ni-based shape memory alloys},
  Progress in materials science 50~(5) (2005) 511--678.

\bibitem{Song_2013}
Y.~Song, X.~Chen, V.~Dabade, T.~W. Shield, R.~D. James, {Enhanced reversibility
  and unusual microstructure of a phase-transforming material}, Nature 502
  (2013) 85.

\bibitem{chluba_2015}
C.~Chluba, W.~Ge, R.~Lima~de Miranda, J.~Strobel, L.~Kienle, E.~Quandt,
  M.~Wuttig, Ultralow-fatigue shape memory alloy films, Science 348~(6238)
  (2015) 1004.

\bibitem{Chen_2013}
X.~Chen, V.~Srivastava, V.~Dabade, R.~D. James, {Study of the cofactor
  conditions: conditions of supercompatibility between phases}, J. Mech. Phys.
  Solids 61 (2013) 2566.

\bibitem{young1993rietveld}
R.~A. Young, {The rietveld method}, Vol.~5, International union of
  crystallography, 1993.

\bibitem{bhattacharya2003}
K.~Bhattacharya, et~al., {Microstructure of martensite: why it forms and how it
  gives rise to the shape-memory effect}, Vol.~2, Oxford University Press,
  2003.

\bibitem{Otsuka_1999}
K.~Otsuka, C.~M. Wayman, Shape memory materials, Cambridge University Press,
  1999.

\bibitem{ball1992}
J.~M. Ball, R.~D. James, F.~Smith, Proposed experimental tests of a theory of
  fine microstructure and the two-well problem, Philosophical Transactions of
  the Royal Society of London. Series A: Physical and Engineering Sciences
  338~(1650) (1992) 389--450.

\bibitem{pitteri_1998symm}
M.~Pitteri, {Geometry and symmetry of multilattices}, International Journal of
  Plasticity 14~(1 -- 3) (1998) 139 -- 157.

\bibitem{hahn1983}
T.~Hahn, U.~Shmueli, J.~W. Arthur, International tables for crystallography,
  Vol.~1, Reidel Dordrecht, 1983.

\bibitem{aroyo2006bilbao}
M.~I. Aroyo, J.~M. Perez-Mato, C.~Capillas, E.~Kroumova, S.~Ivantchev,
  G.~Madariaga, A.~Kirov, H.~Wondratschek, Bilbao crystallographic server: I.
  databases and crystallographic computing programs, Zeitschrift f{\"u}r
  Kristallographie-Crystalline Materials 221~(1) (2006) 15--27.

\bibitem{binaryphase2010}
H.~Okamoto, {Phase diagrams for binary alloys}, ASM International 44 (2010).

\bibitem{Chen_2016jmps}
X.~Chen, Y.~Song, N.~Tamura, R.~D. James, Determination of the stretch tensor
  for structural transformations, Journal of the Mechanics and Physics of
  Solids 93 (2016) 34--43.

\bibitem{Koumatos2016}
K.~Koumatos, A.~Muehlemann, {Optimality of general lattice transformations with
  applications to the Bain strain in steel}, Proceedings of the Royal Society
  A: Mathematical, Physical and Engineering Sciences 472~(2188) (2016)
  20150865.

\bibitem{Koumatos2017}
K.~Koumatos, A.~Muehlemann, {A theoretical investigation of orientation
  relationships and transformation strains in steels}, Acta Crystallographica
  Section A 73~(2) (2017) 115--123.

\bibitem{sutou2008ductile}
Y.~Sutou, T.~Omori, R.~Kainuma, K.~Ishida, {Ductile Cu--Al--Mn based shape
  memory alloys: general properties and applications}, Materials Science and
  Technology 24~(8) (2008) 896--901.

\bibitem{wang2002ebsd}
R.~Wang, J.~Gui, X.~Chen, S.~Tan, {EBSD and TEM study of self-accommodating
  martensites in Cu75. 7Al15. 4Mn8. 9 shape memory alloy}, Acta materialia
  50~(7) (2002) 1835--1847.

\bibitem{fornell2017orientation}
J.~Fornell, N.~Tuncer, C.~Schuh, {Orientation dependence in superelastic
  Cu-Al-Mn-Ni micropillars}, Journal of Alloys and Compounds 693 (2017)
  1205--1213.

\bibitem{sutou1999}
Y.~Sutou, R.~Kainuma, K.~Ishida, {Effect of alloying elements on the shape
  memory properties of ductile Cu--Al--Mn alloys}, Materials Science and
  Engineering: A 273 (1999) 375--379.

\bibitem{warlimont1974}
H.~Warlimont, L.~Delaey, {Martensitic Transformations in Copper- Sliver- and
  Gold-based Alloys}, Vol.~18, Pergamon, 1974.

\bibitem{sutou2002lattparam}
Y.~Sutou, T.~Omori, R.~Kainuma, N.~Ono, K.~Ishida, {Enhancement of
  Superelasticity in Cu-Al-Mn-Ni Shape- Memory Alloys by Texture Control},
  Metall. Mater. Trans. A 33A (2002) 2817 -- 2824.

\bibitem{kainuma1996}
R.~Kainuma, S.~Takahashi, K.~Ishida, {Thermoelastic martensite and shape memory
  effect in ductile Cu-Al-Mn alloys}, Metallurgical and Materials Transactions
  A 27~(8) (1996) 2187 -- 2195.

\bibitem{sutou2004}
Y.~Sutou, T.~Omori, J.~Wang, R.~Kainuma, K.~Ishida, {Characteristics of
  Cu--Al--Mn-based shape memory alloys and their applications}, Materials
  Science and Engineering: A 378~(1 -- 2) (2004) 278 -- 282.

\bibitem{Obrado_1997}
E.~Obrad\'o, L.~Ma\~nosa, A.~Planes, {Stability of the bcc phase of Cu-Al-Mn
  shape-memory alloys}, Physical Review B 56~(1) (1997) 20--23.

\bibitem{Mallik_2008}
U.~S. Mallik, V.~Sampath, {Effect of alloying on microstructure and shape
  memory characteristics of Cu--Al--Mn shape memory alloys}, Materials Science
  and Engineering: A 481--482 (2008) 680--683.

\bibitem{Tamura_2014book}
N.~Tamura, {XMAS: A Versatile Tool for Analyzing Synchrotron X-ray
  Microdiffraction Data}, Imperial College Press, London, 2014.

\bibitem{Chen_2016APL}
X.~Chen, N.~Tamura, A.~MacDowell, R.~D. James, {In-situ characterization of
  highly reversible phase transformation by synchrotron X-ray Laue
  microdiffraction}, Applied Physics Letters 108 (2016) 211902.

\bibitem{Tilley_book}
R.~Tilley, {Crystals and Crystal Structures}, John Wiley \& sons Inc., 2006.

\bibitem{cryst_wyckoff}
Bilbao crystallography server.

\end{thebibliography}

\end{document}